\documentclass[sigconf]{acmart}
\def\BibTeX{{\rm B\kern-.05em{\sc i\kern-.025em b}\kern-.08emT\kern-.1667em\lower.7ex\hbox{E}\kern-.125emX}}
\usepackage[justification=justified,skip=5pt]{caption}

\usepackage{enumitem}
\usepackage{graphicx}
\usepackage{multirow}
\usepackage{bbding}
\usepackage{amsfonts,amssymb}
\usepackage{bm}
\usepackage{booktabs}
\usepackage{url}
\usepackage{algorithm}
\usepackage{algorithmic}
\usepackage{booktabs}

\def\eqref#1{Eq.(\ref{#1})}
\newcommand{\y}{\bm{y}}

\newcommand{\bx}{\bm{X}}

\newcommand{\bw}{\bm{W}}

\copyrightyear{2018}
\acmYear{2018}
\setcopyright{acmlicensed}
\acmConference[Woodstock '18]{Woodstock '18: ACM Symposium on Neural Gaze Detection}{June 03--05, 2018}{Woodstock, NY}
\acmBooktitle{Woodstock '18: ACM Symposium on Neural Gaze Detection, June 03--05, 2018, Woodstock, NY}
\acmPrice{15.00}
\acmDOI{10.1145/1122445.1122456}
\acmISBN{978-1-4503-9999-9/18/06}

\begin{document}

%
% The "title" command has an optional parameter, allowing the author to define a "short title" to be used in page headers.
\title{Privileged Features Distillation at Taobao Recommendations}

\author{Chen Xu*, Quan Li*, Junfeng Ge, Jinyang Gao, Xiaoyong Yang, Changhua Pei, Fei Sun, Jian Wu, Hanxiao Sun, and Wenwu Ou}
\thanks{*Both authors contributed equally to this work}
\affiliation{%
	\institution{Alibaba Group, Beijing, China}
	\city{}
	\country{}
}
\email{{chaos.xc, liquan.quanli, beili.gjf, jinyang.gjy, xiaoyong.yxy, changhua.pch, ofey.sf,joshuawu.wujian}@alibaba-inc.com}

\renewcommand{\shortauthors}{Xu et al.}

\begin{abstract}
Features play an important role in the prediction tasks of e-commerce recommendations. To guarantee the consistency of off-line training  and on-line serving, we usually utilize the same features that are both available. However, the consistency in turn neglects some discriminative features. For example, when estimating the conversion rate (CVR), i.e., the probability that a user would purchase the item if she clicked it, features like dwell time on the item detailed page are informative. However,  CVR prediction should be conducted for on-line ranking before the click happens. Thus we cannot get such post-event features during serving.

We define the features that are discriminative but only available during training as the privileged features. Inspired by the distillation techniques which bridge the gap between training and inference, in this work, we propose privileged features distillation (PFD).  We train two models, i.e., a student model that is the same as the original one and a teacher model that additionally utilizes the privileged features. Knowledge distilled from the more accurate teacher is transferred to the student, which helps to improve its prediction accuracy. During serving, only the student part is extracted and it relies on no privileged features.  We conduct experiments on two fundamental prediction tasks at Taobao recommendations, i.e., click-through rate (CTR) at coarse-grained ranking and CVR at fine-grained ranking. By distilling the interacted features that are prohibited during serving for CTR and the post-event features for CVR, we achieve significant improvements over their strong baselines. During the on-line A/B tests, the click metric is improved  by $\mathbf{+5.0\%}$ in the CTR task.  And the conversion metric is improved by $\mathbf{+2.3\%}$ in the CVR task. Besides, by addressing several issues of training PFD,  we obtain comparable training speed as the baselines without any distillation.
\end{abstract}

\begin{CCSXML}
<ccs2012>
<concept>
<concept_id>10002951.10003317</concept_id>
<concept_desc>Information systems~Information retrieval</concept_desc>
<concept_significance>500</concept_significance>
</concept>
<concept>
<concept_id>10010147.10010257.10010293.10010294</concept_id>
<concept_desc>Computing methodologies~Neural networks</concept_desc>
<concept_significance>500</concept_significance>
</concept>
</ccs2012>
\end{CCSXML}

\ccsdesc[500]{Information systems~Information retrieval}
\ccsdesc[500]{Computing methodologies~Neural networks}

%
% Keywords. The author(s) should pick words that accurately describe the work being
% presented. Separate the keywords with commas.
\keywords{Privileged Features, Distillation, E-commerce Recommendations, CTR, CVR}

%
% A "teaser" image appears between the author and affiliation information and the body 
% of the document, and typically spans the page. 
%   \includegraphics[width=\textwidth]{sampleteaser.pdf}
%   \Description{Enjoying the baseball game from the third-base seats. Ichiro Suzuki preparing to bat.}

%
% This command processes the author and affiliation and title information and builds
% the first part of the formatted document.
\maketitle

\section{Introduction}
In recent years, deep neural networks (DNNs) \cite{wide&deep, youtube, deepfm, xdeepfm, din, dupn} have achieved very promising results in the prediction tasks of recommendations. However, most of these works focus on the model aspect. While there are limited works  except \cite{wide&deep, youtube} paid attention to the feature aspect in the input, which essentially determine the upper-bound of the model performance. In this work, we also focus on the feature aspect, especially the features in e-commerce recommendations. 

To ensure the consistency of off-line training  and on-line serving, we usually use the same features that are both available in the two environments in real applications. However, a bunch of discriminative features, which are only available at training time, are thus ignored. Taking  conversion rate (CVR) prediction in e-commerce recommendations as an example, here we aim to estimate the probability that the user would purchase the item if she clicked it. Features describing user behaviors in the clicked detail page, e.g.,  the dwell time on the whole page, can be rather helpful. However, these features cannot be utilized for on-line CVR prediction in recommendations, because it has to be done before any click happens. Although such post-event features can indeed be recorded for off-line training. In consistent with the learning using privileged information \cite{privilegedinformationlearning,A_new_learning}, here we define the features that are discriminative for prediction tasks but only available at training time, as the \textit{privileged features}.
	
A straightforward way to utilize the privileged features is multi-task learning \cite{mtl}, i.e., predicting each feature with an additional task.  However, in the multi-task learning, each task does not necessarily satisfy a no-harm guarantee (i.e., privileged features can harm the learning of the original model). More importantly, the no-harm guarantee will very likely be violated since estimating the privileged features might be even more challenging than the original problem \cite{privileged_droput}. From the practical point of view, when using dozens of privileged features at once, it would be a challenge to tune all the tasks. 

\begin{figure}[t!]
	\centering
	\includegraphics[width=0.95\linewidth]{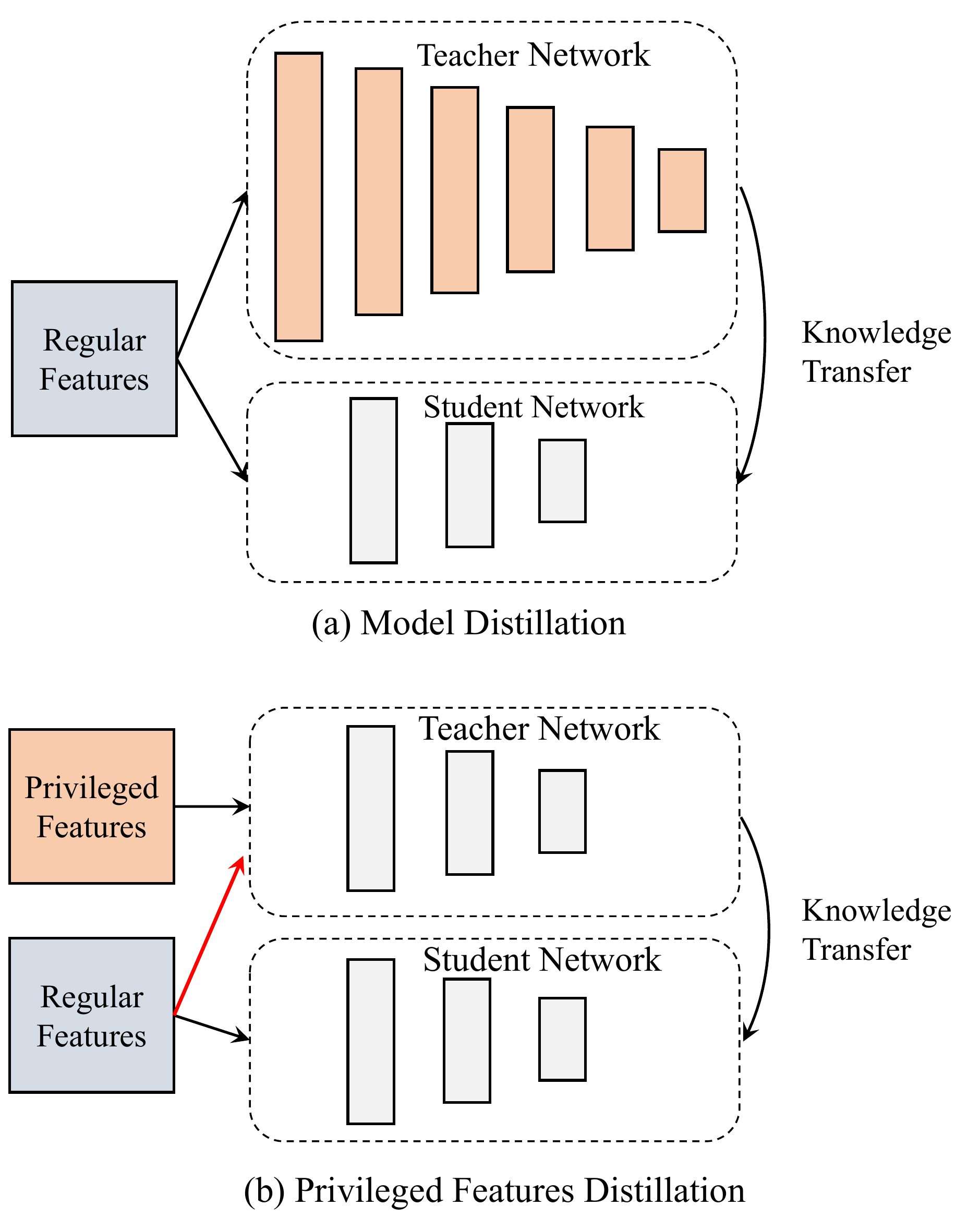}
	{\vspace{-0.5mm}}
	\caption{Illustration of model distillation (MD) \cite{distill}  and privileged features distillation (PFD) proposed in this work. In MD, the knowledge is distilled from the more complex model. While in PFD, the knowledge is distilled from \textit{both} the privileged \textit{and} the regular features. PFD also differs from the original learning using privileged information (LUPI) \cite{dis_privileged}, where the teacher \textit{only} processes the privileged features.} \label{MD&PFD1}
	{\vspace{-0.5mm}}
\end{figure}

Inspired by learning using privileged information (LUPI) \cite{dis_privileged}, here we propose privileged features distillation (\textbf{PFD}) to take advantage of such features.  We train two models, i.e., a student and a teacher model.  The student model is the same as the original one, which processes the  features that are both available for off-line training and on-line serving. The teacher model processes all features, which include the privileged ones. Knowledge distilled from the teacher, i.e., the soft labels in this work, is then used to supervise the training of the student in addition to the original hard labels, i.e., $\{0,1\}$, which additionally improves its performance. During on-line serving, only the student part is extracted, which relies on no privileged features as the input and guarantees the consistency with training. Compared with MTL, PFD mainly has two advantages. On the one hand, the privileged features are combined in a more appropriate way for the prediction task. Generally, adding more privileged features will lead to more accurate predictions. On the other hand, PFD only introduces one extra distillation loss no matter what the number of privileged features is, which is much easier to balance.

PFD is different from the commonly used model distillation (MD) \cite{distill,modelcompress}. In MD, both the teacher and the student process the same inputs. And the teacher uses models with more capacity than the student. For example,  the teachers can use deeper networks to instruct the shallower students \cite{rocket,dis_translation}. Whereas in PFD, the teacher and the student use the same models but differ in the inputs. PFD is also different from the original LUPI \cite{dis_privileged}, where the teacher network in PFD additionally processes the regular features. Figure \ref{MD&PFD1} gives an illustration on the differences. 

In this work, we apply PFD to Taobao recommendations. We conduct experiments on two fundamental prediction tasks by utilizing the corresponding privileged features. The contributions of this paper are four-fold:
\begin{itemize}
\item We identify the privileged features existing at Taobao recommendations and propose PFD to leverage them. Compared with MTL to predict each privileged feature independently, PFD unifies all of them and provides a one-stop solution.

\item Different from the traditional LUPI, the teacher of PFD additionally utilizes the regular features, which instructs the student much better. PFD is in complementary to MD. By combining both of them, i.e., PFD+MD, we can achieve further improvements.

\item We train the teacher and the student synchronously by sharing common input components. Compared to training them asynchronously with independent components as done in tradition, such training manner achieves even or better performance, while the cost of time is much reduced. Thus the technique is  adoptable in online learning, where real-time computation is in high demand.

\item We conduct experiments on two fundamental prediction tasks at Taobao recommendations, i.e., CTR prediction at coarse-grained ranking and CVR prediction at fine-grained ranking.  By distilling the \textit{interacted features} that are prohibited due to efficiency requirement for CTR at coarse-grained ranking and the \textit{post-event features} for CVR as introduced above, we achieve significant  improvements over their strong baselines. During the on-line A/B tests, the click metric is improved  by $\mathbf{+5.0\%}$ in the CTR task.  And the conversion metric is improved by $\mathbf{+2.3\%}$ in the CVR task.

\end{itemize}

\section{Related Distillation Techniques}
Before giving detailed description of our PFD, we will firstly introduce the distillation techniques \cite{distill,modelcompress}. Overall, the techniques are aiming to help the non-convex student models to train better. For model distillation, we can typically write the objective function as follows: 
\begin{equation} \label{model_distill}
\min_{\bw_s}~~ (1-\lambda ) * L_s\left(\y, f_s(\bx; \bw_s) \right) + \lambda * L_d \left(f_t(\bx; \bw_t), f_s(\bx; \bw_s)\right),
\end{equation}
where $f_t$ and $f_s$ are the teacher model and the student model, respectively. $L_s$ denotes the student pure loss with the known hard labels $\y$ and  $L_d$ denotes its loss with the soft labels produced by the teacher. $\lambda \in [0,1]$ is the hyper-parameter to balance the two losses. Compared with the original function that minimizes $L_s$ alone, we are expecting that the additional loss $L_{d}$ in \eqref{model_distill} will help to train $\bw_s$ better by distilling the knowledge from the teacher.  In the work of \cite{regularize_distill}, Pereyra et. al. regard the distillation loss as regularization on the student model.  When training $f_s$ alone by minimizing $L_s$, it is prone to get overconfident predictions, which overfit the training set \cite{noise}. By adding the distillation loss, $f_s$ will also approximate the soft predictions from $f_t$. By softening the outputs, $f_s$ is more likely to achieve better generalization performance. 

Typically, the teacher model is more powerful  than the student model. Teachers can be the ensembles of several models \cite{modelcompress,distill,dis_mutulearn}, or DNNs with more neurons \cite{dis_rank}, more layers \cite{rocket,dis_translation}, or even broader numerical precisions \cite{dis_quant} than students. There are also some exceptions, e.g., in the work of \cite{dis_distribute}, both of the two models are using the same structure and learned from each other, with difference only in the initialization and the orders to process the training data. 

As indicated in \eqref{model_distill}, the parameter $\bw_t$ of the teacher is fixed across the minimization. We can generally divide the distillation technique into two steps:  firstly train the teacher with the known labels $\y$, then train the student by minimizing \eqref{model_distill}.  In some applications, the models could take rather long time to converge, thus it is impractical to wait for the teacher to be ready as \eqref{model_distill}. Instead,  some works try to train the teacher and the student synchronously \cite{dis_distribute,rocket,dis_mutulearn}. Besides distilling from the final output as \eqref{model_distill}, it is possible to distill from the middle layer, e.g., Romero et al. \cite{dis_featuremap}  try to distill the intermediate feature maps, which help to train a deeper and thinner network. 

In addition to distilling knowledge from more complex models,  Lopez-Paz et al. \cite{dis_privileged} propose to distill knowledge from privileged information $\bx^*$, which is also known as learning using privileged information (LUPI). The loss function then becomes:
\begin{equation} \label{privileged_information_distill}
\min_{\bw_s}~~ (1- \lambda) * L_s\left( \y, f(\bx; \bw_s) \right) + \lambda * L_d \left(f(\bx^*; \bw_t), f(\bx; \bw_s)\right).
\end{equation}
In the work of \cite{kdgan}, Wang et al. apply LUPI to image tag recommendation. Besides the teacher and the student, they additionally learn a discriminator, which  ensures the student to learn the true data distribution at the equilibrium faster. Chen et al. apply LUPI to review-based recommendation. They also utilize adversarial training to select informative reviews. Although achieving better performance, both of the works are only validated on relatively small datasets. It still remains unknown whether these techniques can reach equilibrium of the min-max game in industry-scale datasets.

\begin{figure}[t!]
	\centering
	\includegraphics[width=0.97\linewidth]{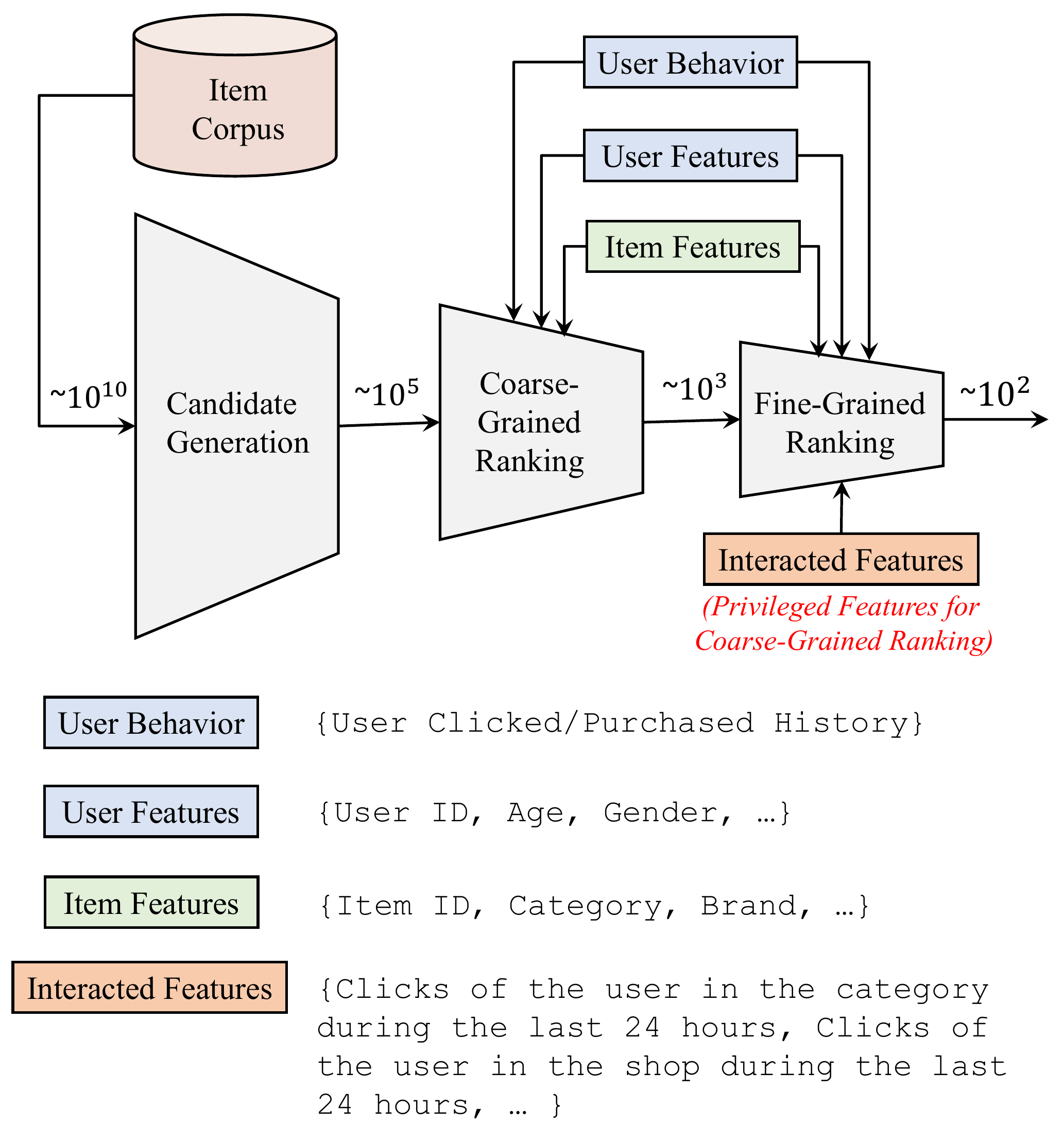}
	\vspace{-0.5mm}
	\caption{Overview of Taobao recommendations. We adopt a cascaded learning framework to select/rank items. At coarse-grained ranking, the interacted features, although being discriminative, are prohibited as they greatly increase the latency at serving.  Some representative features are illustrated in the lower part.}
	\label{cascadedlearning}
	\vspace{-0.5mm}
\end{figure}
\section{Privileged Features at Taobao Recommendations}
To better understand the privileged features exploited in this work,  we firstly give an overview of Taobao recommendations in Figure \ref{cascadedlearning}. As usually done in industry recommendations \cite{youtube, xinai_cascade}, we adopt the cascaded learning framework. There are overall three stages to select/rank the items before presenting to the user, i.e., candidate generation, coarse-grained ranking, and fine-grained ranking. To make a trade-off between efficiency and accuracy, more complex and effective model is adopted as the cascaded stage goes forward, with the expense of higher latency to scoring the items. In the candidate generation stage, we choose around $10^5$  items that are most likely to be clicked or purchased by a user from the huge-scale corpus. Generally, the candidate generation is mixed from several sources, e.g., collaborative filtering \cite{etrac}, the DNN models \cite{youtube}, etc. After the candidate generation, we adopt two stages for ranking, where PFD is applied in this work.

\textbf{In the coarse-grained ranking stage}, we are mainly to  estimate the CTRs of all items selected by the candidate generation stage, which are then used to select the top-$k$ highest ranked items for the next stage.  The inputs of the prediction model mainly consist of three parts. The first part consists of the user behavior, which records the history of her clicked/purchased items.  As the user behavior is in sequential, RNNs \cite{lstm,gru4rec} or self-attention\cite{attentionall,sasrec} is usually adopted to model the user's long short-term interests.  The second part consists of the user features, e.g., user id,  age, gender, etc. and the third part consists of the item features, e.g., item id, category, brand, etc. Across this work, all features are transformed into categorical type and we learn an embedding for each one\footnote{Numerical features are discretized with pre-defined boundaries.}.  

In the coarse-grained ranking stage, the complexity of the prediction model is strictly restricted, in order to  grade tens of thousands of candidates in milliseconds. Here we utilize the inner product model \cite{dssm} to measure the item scores:
 \begin{equation} \label{inner_product_match}
 f\left(\bx^u, \bx^i; \bw^u, \bw^i \right)  \triangleq  \left< \mathbf{\Phi}_{{\bw}^u}\left(\bx^u\right), \mathbf{\Phi}_{\bw^i} \left(\bx^i\right) \right>,
 \end{equation}
where the superscript $u$ and $i$ denote the user and item, respectively.  $\bx^u$ denotes a combination of user behavior and user features. $\mathbf{\Phi}_{\bw}(\cdot)$ represents the non-linear mapping with learned parameter $\bw$.  $\langle \cdot,\cdot \rangle$ is the inner product operation. As the user side and the item side are separated in \eqref{inner_product_match}, during serving, we can compute the mappings $\mathbf{\Phi}_{\bw^i} \left(\cdot\right)$ of all items off-line in advance\footnote{In order to capture the real-time user preference,  the user mappings cannot be stored.}. When a request comes, we only need to execute \textit{one} forward pass to get the user mapping $\Phi_{\bw^u}\left(\bx^u\right)$ and compute its inner product with all candidates, which is extremely efficient. For more details, see the illustration in Figure \ref{inner_product_serving}.

As shown in Figure \ref{cascadedlearning}, the coarse-grained ranking does not utilize any interacted features, e.g., clicks of the user in the item category during the last $24$ hours, clicks of the user in the item shop during the last $24$ hours, etc.  As verified by the experiments below, adding these features can largely enhance the prediction performance. However, it in turn greatly increases the latency during serving, since the interacted features are depending on the user and the specific item. In other words, the features vary with different items or users. If putting them either at the item or the user side of \eqref{inner_product_match}, the inference of the mappings $\Phi_{\bw}\left( \cdot \right)$ need to be executed as many times as the number of candidates, i.e., $10^5$ here. Generally,  the non-linear mapping $\Phi_{\bw}\left( \cdot \right)$ costs several orders more computation than the simple inner product operation.  It is thus unpractical to use the interacted features during serving. Here we regard them as the \textit{privileged features} for CTR prediction at coarse-grained ranking.

\begin{figure}[t!]
	\centering
	\includegraphics[width=0.95\linewidth]{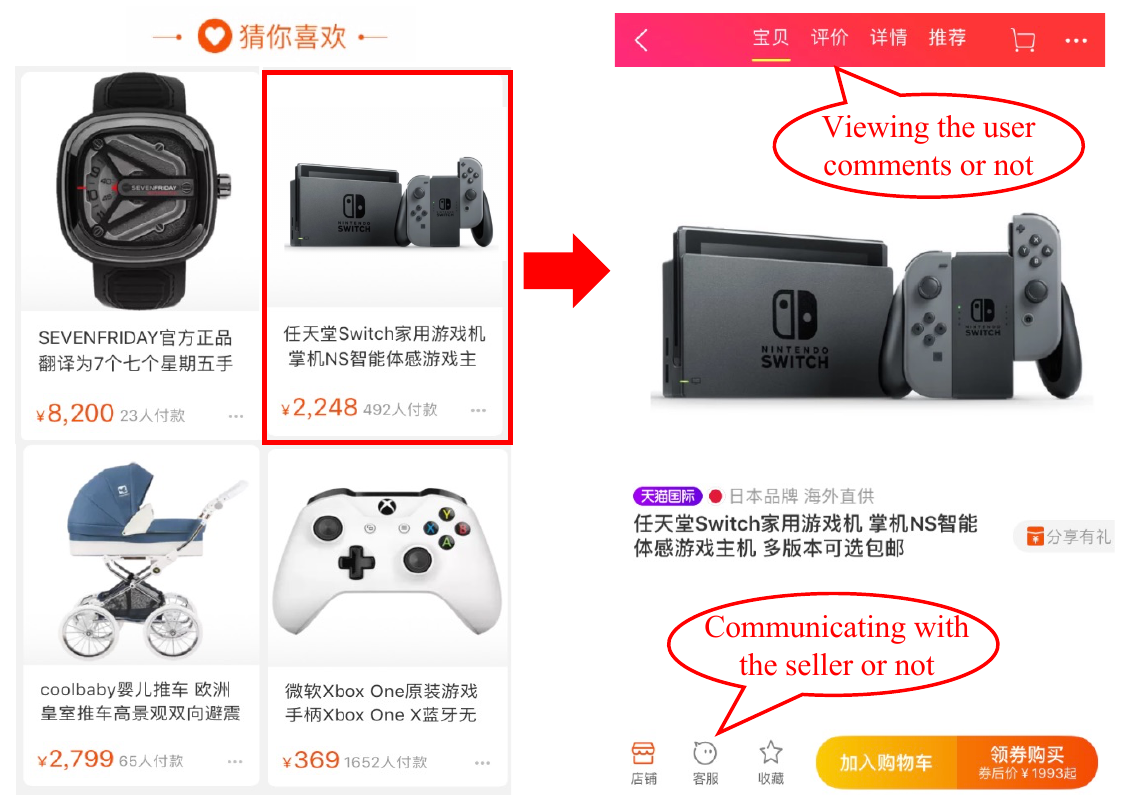} 
	\vspace{-0.5mm}
	\caption{Illustrative features describing the user behavior in the detailed page of the clicked item. Including the dwell time that is not shown, these features are rather informative for CVR prediction. However, during serving, we need to use CVR to rank all candidate items as shown in left sub-figure before any item being clicked. We thus denote these features as the privileged features for CVR prediction.}
	\label{cvr_illustration}
	\vspace{-0.5mm}
\end{figure}

\textbf{In the fine-grained ranking stage}, besides estimating the CTR as done in the coarse-grained ranking, we need to estimate the CVR for all candidates, i.e., the probability that the user would purchase the item if she clicked it. In e-commerce recommendations, the main aim is to maximize the Gross Merchandise Volume (GMV), which can be decomposed into CTR $\times$ CVR $\times$ Price. Once estimating the CTR and CVR for all items, we can rank them by the expected GMVs to maximize them. Under the definition of CVR, it is obvious that user behaviors on the detailed page of the clicked item, e.g., dwell time, whether viewing the comment or not, whether communicating with the seller or not, etc., can be rather helpful for the prediction. However, CVR needs to be estimated for ranking before any future click happens. Features describing user behaviors on the detailed page are not available during inference. Here we thus denote these features as the \textit{privileged features} for CVR prediction. To better understand that, we give an illustration in Figure \ref{cvr_illustration}.

\section{Privileged Feature Distillation}\label{section4}

\renewcommand{\algorithmicrequire}{\textbf{Input:}}
\renewcommand{\algorithmicensure}{\textbf{Output:}}
\begin{algorithm}[tb]
	\caption{Minimizing the student, distillation, and teacher loss in \eqref{feature_distill_teacher} synchronously with SGD.}  
	\label{algorithm1}
	%\textbf{Output}: Your algorithm's output
	\begin{algorithmic}[1] %[1] enables line numbers
		\REQUIRE Hyper-parameter $\lambda$, swapping step $k$, and learning rate $\eta$
		\STATE Initialize $(\bw_s, \bw_t)$ and let $i=0$. 
		\WHILE{not converged}
		\STATE Get training data $\left(\y, \bx,\bx^* \right).$
		\IF {$i < k$}
		\STATE  $\bw_s = \bw_s - \eta \nabla_{\bw_s}L_s.$ 
		\ELSE
		\STATE $\bw_s = \bw_s - \eta \nabla_{\bw_s}\{(1- \lambda) *L_s +
		\lambda * L_d  \}.$
		\ENDIF		
		\\$\bw_t = \bw_t - \eta \nabla_{\bw_t}L_t.$ 
		//\small{\mbox{\tt No distillation loss} $L_d$}
		\STATE Update $i =  i+1$
		\ENDWHILE
		\ENSURE $(\bw_s, \bw_t)$
	\end{algorithmic}
\end{algorithm}

In the original LUPI as \eqref{privileged_information_distill}, the teacher relies on the privileged information $\bx^*$.  Although being informative,  the privileged features in this work only partially describe the user's preference. The performance of which using these features can even be inferior to that of which using regular features. Besides, the predictions based on privileged features can sometimes be misleading. For example, it generally takes more time for customers to decide on expensive items, while the conversion rate of these items is rather low.  When conducting CVR estimation, the teacher of LUPI makes predictions relying on the privileged features, e.g., dwell-time,  but not considering the regular features, e.g., item price, which may result in false positive predictions on the expensive items. To alleviate that, we additionally feed the regular features to the teacher model. The  original function of \eqref{privileged_information_distill} is thus modified as follows:
 \begin{equation} \label{feature_distill}
\resizebox{.92\linewidth}{!}{$
	\displaystyle
\min_{\bw_s}~~ (1 - \lambda) * L_s\left(\y, f(\bx; \bw_s) \right) + \lambda * L_d \left(f(\bx, \bx^*; \bw_t), f(\bx; \bw_s)\right).
$}
\end{equation}
Generally, adding more information, e.g., more features, will lead to more accurate predictions. The teacher $f(\bx, \bx^*; \bw_t)$ here is thus expected to be stronger than the student $f(\bx; \bw_s)$ or the teacher of LUPI $f(\bx^*; \bw_t)$. In the above scenario, by taking both the privileged and the regular features into consideration, the dwell time feature instead can be used to distinguish the extent of preference on different expensive items. The teacher is thus more knowledgeable to instruct the student rather than mislead it. As verified by the experiments below, adding regular features to the teacher is non-trivial and it greatly improves the performance of LUPI. Thereafter we denote this technique as PFD to distinguish it from LUPI.
 
As \eqref{feature_distill} indicates, the teacher $f(\bx, \bx^*; \bw_t)$ is trained in advance. However, it takes a long time to train the teacher model alone in our applications. It is thus quite unpractical to apply distillation as \eqref{feature_distill}. A more plausible way is to train the teacher and  the student synchronously as done in \cite{dis_distribute,rocket,dis_mutulearn}. The objective function is then modified as follows:
\begin{equation} \label{feature_distill_teacher}
\resizebox{.92\linewidth}{!}{$
	\displaystyle
\begin{split}
\min_{\bw_s, \bw_t}~~ &(1 - \lambda) * L_s\left( \y, f(\bx; \bw_s) \right) + \lambda * L_d \left(f(\bx, \bx^*; \bw_t), f(\bx; \bw_s)\right) \\&+ L_t\left( \y, f(\bx,\bx^*; \bw_t) \right).
\end{split}
$}
\end{equation}

\begin{figure*}[ht!]
	\centering
	\includegraphics[width=0.99\linewidth]{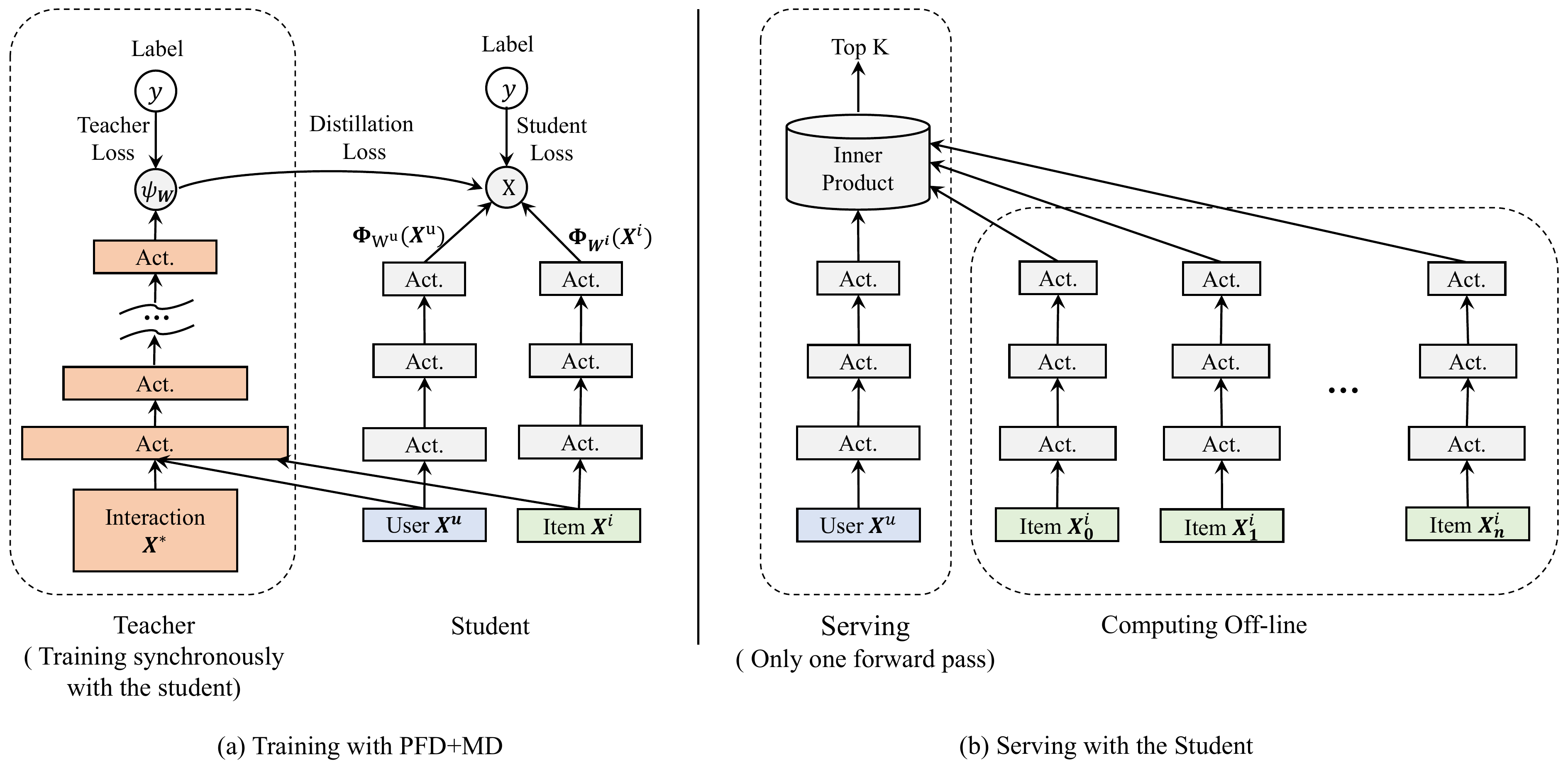} 
	{\vspace{-0.5mm}}
	\caption{Illustration of training the inner-product model with PFD+MD and (b) its deployment during serving.  At the training time, the privileged features, i.e., interaction $\bx^*$ between $\bx^u$ and $\bx^i$, and the more complex DNN model together form a strong teacher to instruct the student.  During serving,  we compute the mappings $\mathbf{\Phi}_{\bw^i} \left(\cdot\right)$ of all items off-line in advance. When a request comes, we only need to execute one forward pass to derive the user mapping $\mathbf{\Phi}_{\bw^u} \left(\bx^u\right)$.   }
	\label{inner_product_serving}
	{\vspace{-0.5mm}}
\end{figure*}

Although saving time, synchronous training  can be un-stable. In the early stage when the teacher is not well-trained, the distillation loss $L_d$ may distract the student and slow the training down. Here we alleviate it by adopting a warm up scheme. We set $\lambda$ of \eqref{feature_distill_teacher} to $0$ in the early stage and fix it to the pre-defined value thereafter, with the swapping step being a hyper-parameter. In our huge scale datasets, we find that such simple scheme works well.  Different from mutual learning \cite{dis_mutulearn},  we only enable the student to learn from the teacher here. Otherwise the teacher will co-adapt with the student, which degenerates its performance.  When computing the gradient with respect to the teacher parameters $\bw_t$, we thus omit the distillation loss $L_d$. The update with SGD is illustrated in Algorithm \ref{algorithm1}. 

Across this work, all models are trained in the parameter sever systems \cite{distbelieve}, where all parameters are stored in the servers and most computation is executed in the workers. The training speed is mainly depending on the computation load in the workers and the communication volume between the workers and the servers. As indicated in \eqref{feature_distill_teacher}, we  train the teacher and the student together. The number of parameters and the computation are roughly doubled. Training using PFD can thus be much slower than training on the student alone, which is unpractical in industry. Especially for on-line learning where real-time computation is in high demand, adopting distillation can add much burden. Here we alleviate that by sharing all common input components in the teacher and the student.  Since the embeddings of all features occupy most of the storage in the severs\footnote{For the student model alone, all embeddings take up to $150$ Gigabytes.}, the communication volume is almost halved by sharing. The computation can also be reduced by sharing the components of processing the user clicked/purchased behavior, which is known to be costly. As verified by the experiments below,  we can achieve even or better performance by sharing.  Besides, we increase only a little extra time compared to training the student alone, which makes PFD adoptable for online learning. 

\noindent \textbf{Extension: PFD+MD.} As illustrated in Figure~\ref{MD&PFD1}, PFD distills knowledge from the privileged features. In comparison, MD distills knowledge  from the more complex teacher model.  The two distillation techniques are complementary.  A natural extension is to combine them by forming a more accurate teacher to instruct the student. 

In the CTR prediction at coarse-grained ranking, as \eqref{inner_product_match} shows, we use the inner product model to increase the efficiency during serving. In fact, the inner product model can be regarded as the generalized matrix factorization \cite{youtube}. Although we are using \textit{non-linear} mapping $\mathbf{\Phi}_{\bw}(\cdot)$ to transform the user and item inputs, the model capacity is intrinsically limited by the \textit{bi-linear} structure at the inner product operation. DNNs, with the capacity to approximate any function \cite{approximation1, approximation2}, are  considered as a substitution for the inner product model in the teacher. In fact, as proved in Theorem 1 of~\cite{why},  the product operation can be approximated arbitrarily well by a two-layers neural network with only $4$ neurons in the hidden layer. Thus the performance of using DNN is supposed to be lower-bounded by that of using the inner-product model. 

\begin{figure*}[ht!]
	\centering
	\includegraphics[width=0.97\linewidth]{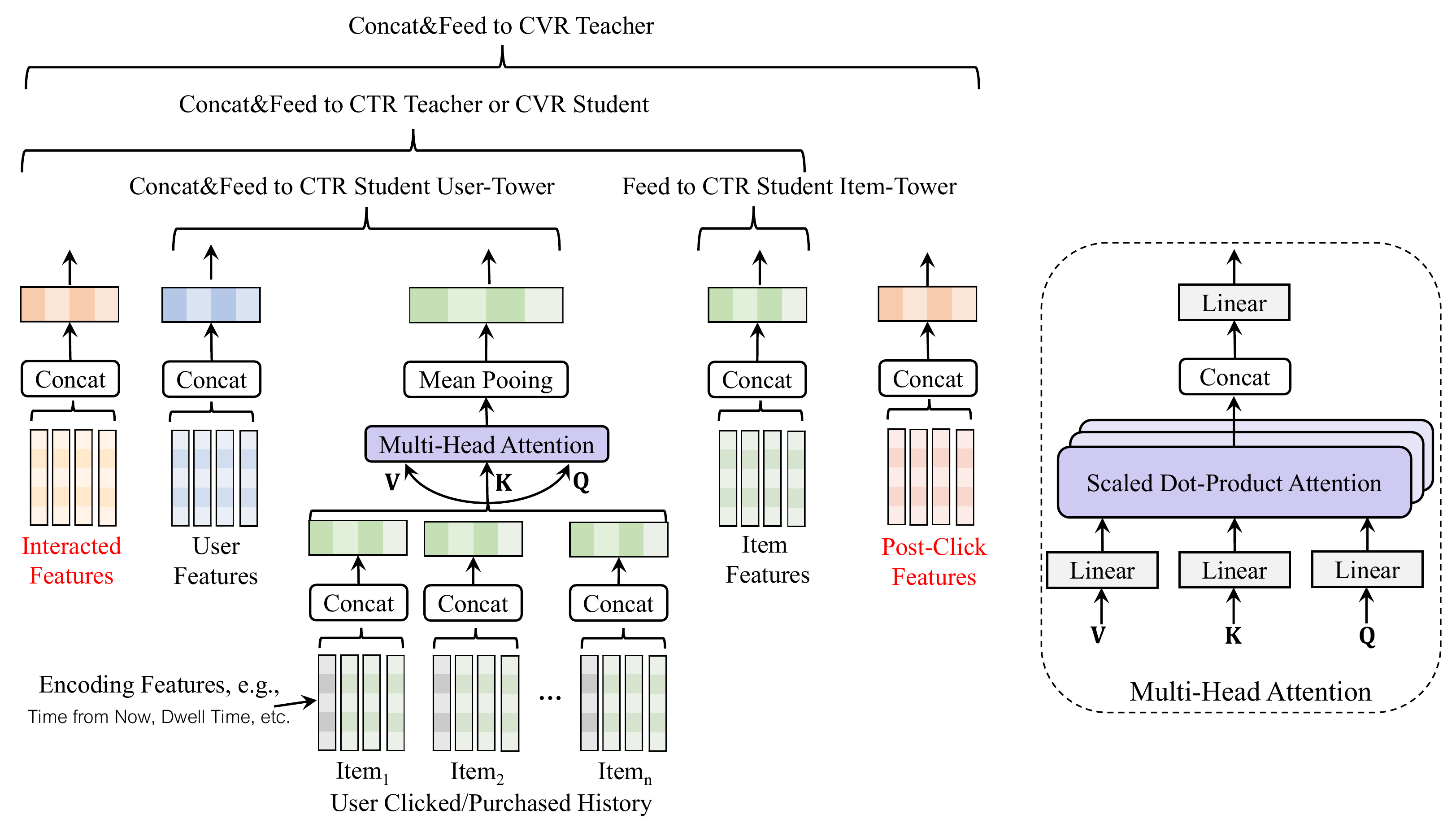} 
	{\vspace{-0.5mm}}
	\caption{Illustration of input components used for CTR at coarse-grained ranking and CVR at fine-grained ranking. We adopt the multi-head self-attention structure \cite{attentionall} to model the user clicked/purchased history. The common input components between the teacher and the student are shared during training. }
	\label{input_components}
	{\vspace{-0.5mm}}
\end{figure*}

In PFD+MD,  we thus adopt the DNN model as the teacher network.  In fact, the teacher model here is the same as the model used for CTR prediction at fine-grained ranking. PFD+MD in this task can be regarded as distilling knowledge from the fine-grained ranking to improve the coarse-grained ranking. For better illustration, we give the whole framework in Figure \ref{inner_product_serving}.  During serving, we extract the student part only, which relies on no privileged features. As the mappings $\mathbf{\Phi}_{\bw^i} \left(\bx^i\right)$ of all items are independent of the users, we compute them off-line in advance. When a request comes, the user mapping $\mathbf{\Phi}_{\bw^u} \left(\bx^u\right)$ is firstly computed.  After that, we compute its inner-product with the mappings of all items produced from the candidate generation stage. The top-$k$ highest scored items are  then chosen and fed to the fine-grained ranking. On the whole, we only execute one forward pass to derive the user mapping and conduct efficient inner product operations between the user and all candidates, which are rather friendly in the aspect of computation.

\section{Experiments}
In this section, we conduct experiments at Taobao recommendations with the aim of answering the following research  questions:
\begin{itemize}
	\item \textbf{RQ1:} What is the performance of PFD on the tasks of CTR at coarse-grained ranking and CVR at fine-grained ranking?
	\item \textbf{RQ2:} Compared to individual PFD, can we achieve additional improvements by combining PFD with MD? 
	\item \textbf{RQ3:} Is PFD sensitive to the hype-parameter $\lambda$ in \eqref{feature_distill_teacher}?
	\item \textbf{RQ4:} What is the effect of training the teacher and student synchronously  by sharing common input components?
\end{itemize}
\subsection{Experimental Settings}
To better understand the network structure, we give an illustration of all input components used in this work in Figure \ref{input_components}. As mentioned earlier, all features are transformed into categorical type and we learn an embedding for each one.  The entities, e.g., user and item, are then represented by the concatenations of their corresponding feature embeddings. Here we adopt the self-attention structure \cite{attentionall} to model the user clicked/purchased history. Let $\mathbf{V}\in \mathbb{R}^{n \times k}$ denote the input values of the sequence, with $n$ being the combined embedding dimension and $k$ being the sequence length. The transformed values $\mathbf{\tilde{V}}$ are then derived as weighted combinations of input $\mathbf{V}$, i.e.,
$\mathbf{\tilde{V}} =  \text{Softmax}\left(\mathbf{Q}^T\mathbf{K}\right)\mathbf{V}$,
where the queries $\mathbf{Q}$ and the keys $\mathbf{K}$ are the same as the values $\mathbf{V}$. To improve the effective resolution of combinations, the multi-head structure is adopted, where  $\mathbf{Q}$, $\mathbf{K}$, and $\mathbf{V}$ are projected linearly into several subspaces and the transformations are executed, respectively.  Note that we also add feed-forward network, skip connections \cite{skipconnect}, layer normalization \cite{layernorm} to the attention mechanism as done in the original work \cite{attentionall}. For simplicity, we omit the details here. The only difference with the original self-attention is that we do not adopt position encodings in the input.  We instead insert several extra features describing the user behavior on the particular item, e.g., the clicked/purchased time from now, the dwell time on the item, etc. Besides encoding the relative positions of the items, the extra features also reflect the importance of the items for future predictions. According to our experiments, adding these features can greatly improve the performance. 
 
After deriving $\mathbf{\tilde{V}}$ with one-layer's self-attention, we adopt the mean pooling operation over the sequence length $k$. Here we set $k$ to $50$, the number of heads to $4$, and the subspace dimension to $32$. As illustrated in Figure \ref{input_components}, the input components are fed to the corresponding teacher and student networks, which consists of several fully-connected layers.  We use LeakyReLU \cite{leakyrelu} as the activation and insert batch normalization \cite{bn} before it.  The models are trained in the parameter servers with the asynchronous Adagrad optimizer \cite{adagrad}.  In the first one million steps, the learning rate is increased linearly to the pre-defined value $0.01$, which is then kept fixed. We set the batch size to $1024$ and the number of epoch to $1$. We train the teacher and  student synchronously by sharing common input components.  Unless stated otherwise, $\lambda$ is set to $0.5$ and the swapping step $k$ in Algorithm \ref{algorithm1} is set to $10^6$.
 
As the labels are in $1$ or $0$, i.e., whether the users clicked/purchased the item or not, we use the log-loss for both the teacher and the student, i.e.,
\begin{equation}
\resizebox{.91\linewidth}{!}{$
	\displaystyle
	L_{t/s} \triangleq  \frac{1}{N} \sum_{i =1}^{N}\left(y_i \mbox{log}\left(1+e^{-f_{t/s,i}} \right) + (1 - y_i)\mbox{log}\left(1+e^{f_{t/s,i}} \right) \right), 
	$}
\end{equation}
where $f_{t/s,i}$ denotes the output of the $i$-th sample from the teacher or student model. For the distillation loss $L_d$,  we use the cross entropy, i.e., by replacing $y_i$ in the above equation with $1/(1 + e^{-f_{t,i}})$. Here we measure the performance of models with the widely-used areas under the curve (AUC) in the next-day held-out data.

\subsection{CTR at Coarse-grained Ranking (RQ1-2)}
\begin{table}
	\centering
	\caption{CTR dataset description at coarse-grained ranking of Taobao recommendations.}
	\vspace{-0.5mm}
	\label{ctr_dataset}
	\setlength{\tabcolsep}{1.5mm}{
		\begin{tabular}{c|cccc}
			\toprule
			Datasets  &   $\#$ Users   &  $\#$Items & $\#$Clicks   & $\#$ Impressions  \\
			\midrule
			\midrule
			$1$ Day &   $9.35 \times 10^{7}$  &  $2.67 \times 10^{7}$    & $5.03 \times 10^{8}$  &  $1.09 \times 10^{10}$  \\
			$10$ Days &$2.88 \times 10^{8}$      & $4.45 \times 10^{7}$   &  $4.57 \times 10^{9}$  & $9.90 \times 10^{10}$  \\
			\bottomrule
	\end{tabular}}
	\vspace{-0.5mm}
\end{table}

\begin{table}
	\centering
	\caption{Testing AUC of different methods for CTR at coarse-grained ranking. We do not include MTL as it is too cumbersome to predict dozens of privileged features. Due to the huge training cost, we only compare PFD+MD with the baseline in the dataset of 10 days. }
	\vspace{-0.5mm}
	\label{rs_auc_tabel1}
	\setlength{\tabcolsep}{3.0mm}{
		\begin{tabular}{l|cc|cc}  
			\toprule
		\multirow{2}{*}{Methods}	  &   \multicolumn{2}{|c|}{Dataset of $1$ Day} &  \multicolumn{2}{c}{Dataset of $10$ Days} \\
			    & Student & Teacher & Student & Teacher \\
			\midrule
			\midrule
			Baseline  &  $0.6625$  & $-$       & $0.7042$ & $-$ \\
			\midrule
			LUPI \cite{dis_privileged}       &   $0.6637$   & $0.6687$   &  $-$       & $-$ \\
			MD \cite{distill}         &   $0.6704$   & $0.6892$   &  $-$       & $-$ \\
			PFD        &   $0.6712$    & $0.6921$   &  $-$       & $-$ \\
			PFD+MD &   $\mathbf{0.6745}$   & $0.7110$     & $\mathbf{0.7160}$   &  $0.7411$ \\ 
			\bottomrule
	\end{tabular}}
\vspace{-0.5mm}
\end{table}

Across this work,  we conduct experiments using the traffic logs of the Guess You Like scenario in the front page of Taobao app. The dataset description of CTR at coarse-grained ranking is summarized in Table \ref{ctr_dataset}.  For the inner-product model, the user mapping $\mathbf{\Phi}_{\bw^u} \left(\cdot\right)$ and the item mapping $\mathbf{\Phi}_{\bw^i} \left(\cdot\right)$ in \eqref{inner_product_match} are formulated as: Input->$512$-> Act.->$256$->Act.->$128$->$\ell_2$-normalize. When executing inner-product between $\mathbf{\Phi}_{\bw^u} \left(\cdot\right)$ and  $\mathbf{\Phi}_{\bw^i} \left(\cdot\right)$, we additionally multiply it with a scalar, i.e., $5$, in compensation for the shrinkage of value after normalization.
 In LUPI \cite{dis_privileged},  MD \cite{distill}, and PFD+MD, the teacher networks use $3$-layers' MLP, with the number of hidden neurons being $512$, $256$, and $128$, respectively. In PFD, we use the inner-product model for both the teacher and the student, where the privileged features are put at the user side of the teacher. The teacher and the student of PFD (and PFD+MD) share all  common input components except user id.
 We do not include MTL as it is too cumbersome to predict dozens of privileged features.

\noindent \textbf{Off-line and on-line performance.}  The testing AUC of all compared distillation techniques are shown in the left part of Table \ref{rs_auc_tabel1}. 
Compared the teacher of PFD with the baseline,  we confirm the effectiveness of the  interacted features.  
By distilling knowledge from these features, we improve the testing AUC from $0.6625$ to $0.6712$. 
Although both utilizing the same privileged features, PFD is superior to LUPI.  
This is because that the teacher of PFD additionally processes the regular features, which results in more  accurate predictions to instruct the student than using privileged features alone as done in LUPI ($0.6921$ v.s. $0.6687$). 
Similarly, we achieve further improvements with PFD+MD by forming a more accurate teacher.
In order to validate whether the superiority of PFD+MD could still holds when training longer with more data. We conduct experiments on the traffic logs of $10$ days. Due to the huge training cost, we only compare PFD+MD with the baseline. As shown in the right part of Table \ref{rs_auc_tabel1},  the student of PFD+MD surpasses the baseline AUC with a margin, i.e., $+0.0118$. We also conduct on-line A/B tests to validate its effectiveness. Compared with the baseline, PFD+MD consistently improves the click metric by $\mathbf{+5.0\%}$. And we have  \textbf{fully deployed} the technique into production.

\noindent \textbf{Cost of directly utilizing interacted features.} As discussed earlier, the interacted features are prohibited for the inner-product model during inference. With such features, we need to compute the mappings $\Phi_{\bw}\left( \cdot \right)$ as many times as the number of candidates, i.e., $10^5$. In contrast, without such features, we only need to compute the mapping once and execute its inner product with all candidates efficiently. Suppose that the input to  $\mathbf{\Phi}_{\bw} \left(\cdot\right)$ has $1024$ dimensions. Theoretically, we need $1024 \times 512+512\times256+256\times128 \approx 6.9 \times 10^5$ fused multiply-add flops to get one mapping. In comparison, executing one inner product operation needs $128$ flops, which is $\sim5400\times$ less.  We also conduct simulated experiments in the personal computer.  We repeat $10^5$ times to simulate the mapping inference and the inner product operation, which totally costs $89.695$s and $0.108$s, respectively.  Executing the mapping is $\sim830\times$ slower than the inner product operation.  

\subsection{CVR  at fine-grained ranking (RQ1-2)}

\begin{table}
	\centering
	\caption{CVR dataset description at fine-grained ranking of Taobao recommendations}
	\vspace{-0.5mm}
	\label{cvr_dataset}
	\setlength{\tabcolsep}{1.5mm}{
		\begin{tabular}{c|cccc}
			\toprule
			Datasets  &   $\#$Users   &  $\#$Items &  $\#$Purchases   & $\#$Clicks \\
			\midrule
			\midrule
			$30$ Days & $2.78 \times 10^{8}$       & $3.74 \times 10^{7} $   &  $6.97 \times 10^{7}$  & $1.40 \times 10^{10}$    \\
			$60$ Days & $3.36 \times 10^{8}$       & $5.27 \times 10^{7} $   &  $1.32 \times 10^{8}$  & $2.71 \times 10^{10}$    \\
			\bottomrule
	\end{tabular}}
\vspace{-0.5mm}
\end{table}

\begin{table}
	\centering
	\caption{Testing AUC  of different methods for CVR prediction at fine-grained ranking.}
	\vspace{-0.5mm}
	\label{cvr_auc_tabel}
	\setlength{\tabcolsep}{3.0mm}{
	\begin{tabular}{l|cc|cc}
		\toprule
		\multirow{2}{*}{Methods}& \multicolumn{2}{c|}{Dataset of $30$days} & \multicolumn{2}{c}{Dataset of $60$days} \\
		\qquad  &   Student & Teacher  & Student & Teacher        \\
		\midrule
		\midrule
		Baseline \qquad     & $0.9040$ & $-$     &   $0.9082$           &   $-$      \\
		MTL \cite{mtl} \qquad & $0.9045$ & $-$     &        $0.9077$       &  $-$     \\
		\midrule
		LUPI \cite{dis_privileged} \qquad 
		& $0.8965$ & $0.9651$          &      $0.9003$          &  $0.9659$     \\
		MD \cite{distill} \qquad & $0.9052$ & $0.9058$      &        $0.9093$       &  $0.9103$  \\
		PFD  \qquad&   $\mathbf{0.9084}$  &   $0.9901$     &          $0.9135$     &   $0.9923$  \\
		PFD+MD  \qquad &  $0.9082$ &  $0.9911$ &        $\mathbf{0.9138}$       &   $0.9929$    \\
		\bottomrule
	\end{tabular}}
\vspace{-0.5mm}
\end{table}

In the above CTR prediction, we use the traffic logs of all impressions. While in the CVR prediction, we extract the traffic logs of all clicks.  The datasets are summarized in Table \ref{cvr_dataset}.  We adopt $3$-layers' MLP for the baseline and all of the student networks, with the number of hidden neurons being $512$, $256$, and $128$, respectively. The teacher networks of PFD and LUPI also have the same structure as their students. In MD and PFD+MD, their teacher networks are expanded to $7$-layers' MLP, with the number of hidden neurons being $8192$, $4096$, $2048$, $1024$, $512$, $256$, and $128$, respectively. We also compare with MTL. Here
we adopt the hard parameter sharing version of MTL \cite{mtl}, where all tasks share first hidden layer and independently predict each task with $2$-layers' MLP. We adopt the mean squared error for the auxiliary continuous regression tasks, e.g., predicting the dwell time, and the log-loss for the auxiliary binary prediction tasks, e.g., predicting whether viewing the user comments or not. The hyper-parameters are chosen empirically from $\{0.01, 0.02, 0.05, 0.1, 0.2, 0.5, 1, 2\}$ with the principle of keeping all auxiliary losses balanced.

\begin{table}
	\centering
	\caption{Students' testing AUC of different hyper-parameters $\lambda$ in the $1$ day's dataset of CTR. The superscript $^+$/$^-$ indicates the highest/lowest  AUC of each method among all chosen hyper-parameters.}
	\vspace{-0.5mm}
	\label{ctr_lambda}
	\begin{tabular}{l|lllll}
		\toprule
		\multicolumn{1}{c|}{$\lambda$\qquad}  &  \multicolumn{1}{c}{$0.1$}  & \multicolumn{1}{c}{$0.3$}  & \multicolumn{1}{c}{$0.5$} & \multicolumn{1}{c}{$0.7$} & \multicolumn{1}{c}{$0.9$}        \\
		\midrule
		\midrule
		LUPI \cite{dis_privileged} \qquad &
		$0.6648^+$&$0.6640$ & $0.6637$&$0.6631$&$0.6624^-$\\
		MD \cite{distill} \qquad &  $0.6695^-$&$0.6697$ & $0.6704$&$0.6706^+$&$0.6700$ \\
		PFD  \qquad&   $0.6711$&$0.6709$ & $0.6712^+$&$0.6700$&$0.6696^-$ \\
		PFD+MD  \qquad &  $0.6741$&$0.6740$ & $0.6745$&$0.6747^+$&$0.6739^-$    \\
		\bottomrule
	\end{tabular}
	\vspace{-0.5mm}
\end{table}

\begin{table}
	\centering
	\caption{Students' testing AUC of different hyper-parameters $\lambda$ in the $30$ days' dataset of CVR.}
	\vspace{-0.5mm}
	\label{cvr_lambda}
	\begin{tabular}{l|lllll}
		\toprule
		\multicolumn{1}{c|}{$\lambda$\qquad}  &  \multicolumn{1}{c}{$0.1$}  & \multicolumn{1}{c}{$0.3$}  & \multicolumn{1}{c}{$0.5$} & \multicolumn{1}{c}{$0.7$} & \multicolumn{1}{c}{$0.9$}        \\
		\midrule
		\midrule
		LUPI \cite{dis_privileged} \qquad 
		&$0.9024^+$& $0.8998$& $0.8965$& $0.8876$& $0.8613^-$    \\
		MD \cite{distill} \qquad & $0.9047^-$& $0.9054^+$& $0.9052$& $0.9050$& $0.9049$ \\
		PFD  \qquad&   $0.9081^-$& $0.9082$& $0.9084^+$& $0.9082$& $0.9082$  \\
		PFD+MD \qquad &  $0.9081$ & $0.9085^+$& $ 0.9082$& $ 0.9083$& $ 0.9080^-$    \\
		\bottomrule
	\end{tabular}
	\vspace{-0.5mm}
\end{table}

\noindent \textbf{Off-line and on-line performance.} The results of all testing methods are shown in Table \ref{cvr_auc_tabel}. Among them, LUPI gets the worst performance,  although the testing AUC of its teacher is rather high. The reason of the inferiority has been discussed in Section \ref{section4}. The experiments also confirm the positive effects of adding regular features to the teacher model, where PFD improves the baseline by $+0.0044$ AUC in the dataset of $30$ days and $+0.0053$ AUC in the dataset of $60$ days, respectively. Compared with PFD,  PFD+MD has no distinct superiority. This is mainly due to that the improvement of MD over the baseline is moderately small.  In practice,  PFD is thus preferred over PFD+MD as costing much less computation resources.  We conduct on-line A/B tests to validate the effectiveness of PFD. Compared with the baseline, PFD steadily improves the conversion metric by $\mathbf{+2.3\%}$ over a long period of time.

\subsection{Ablation Study (RQ3-4)}

\textbf{Sensitivity of Hyper-parameter.} In the above experiments, we fix the the hyper-parameter $\lambda$ at $0.5$ for all distillation techniques. Here we test the sensitivity of $\lambda$. The results of CTR prediction are shown in Table \ref{ctr_lambda}. Most of the distillation techniques surpass the baseline, i.e., $0.6625$, among all chosen hyper-parameters, except LUPI with $\lambda= 0.9$.  MD, PFD, and PFD+MD are all robust to varying $\lambda$.  Even their worst results improve the baseline by margins. We also conduct experiments in the CVR dataset. As shown in Table \ref{cvr_lambda}, LUPI narrows the gap with the baseline, i.e., $0.9040$ as $\lambda$ decreases, while it is still inferior to the baseline.  MD, PFD, and PFD+MD are again robust to varying $\lambda$  in this huge-scale dataset. 

\begin{table}
	\centering
	\caption{Effects of training PFD+MD in different manners in the $1$ day's dataset of CTR. Ind\&Async denotes that the teacher and the student are trained asynchronously with independent input components. Share\&Sync denotes that the teacher and the student are trained synchronously with  shared common input components. The superscript $^*$ means that all common input components are shared except user id. We also record the wall-clock time in hours in the forth column.}
	\vspace{-0.5mm}
	\label{effect_of_ctr}
	\setlength{\tabcolsep}{2.5mm}{
		\begin{tabular}{l|cc|cc}
			\toprule
			\qquad  &   Student & Teacher  & Time & Relative  \\
			\midrule
			\midrule
			Baseline  &
			$0.6625$ & $-$          &      $9.24$ h          &  $0\%$     \\
			\midrule
			Ind\&Async & 
			$0.6751$ & $0.7112$          &  $18.43$ h          &  $+99.5\%$     \\
			Ind\&Sync&   
			$0.6748$  &   $0.7112$     &          $14.32$ h    &   $+55.0\%$  \\
			Share\&Sync  &  
			$0.6717$ &  $0.7108$ &       $9.51$ h       &  $+2.9\%$     \\
			Share$^*$\&Sync  &  
			$0.6745$ &  $0.7110$ &        $10.29$ h       &   $+11.4\%$    \\
			\bottomrule
	\end{tabular}}
	\vspace{-0.5mm}
\end{table}

\begin{table}
	\centering
	\caption{Effects of training PFD+MD in different manners in the $30$ days' dataset of CVR. The row with superscript $^{\dagger}$ is the result of PFD.}
	\label{effect_of_cvr}
	\vspace{-0.5mm}
	\setlength{\tabcolsep}{2.5mm}{
		\begin{tabular}{l|cc|cc}
			\toprule
			\qquad  &   Student & Teacher  & Time & Relative  \\
			\midrule
			\midrule
			Baseline  &
			$0.9040$ & $-$          &      $12.22$ h          &  $0\%$     \\
			\midrule
			Ind\&Async & 
			$0.9067$ & $0.9887$          &  $26.85$ h          &  $+119.7\%$     \\
			Ind\&Sync&   
			$0.9069$  &   $0.9887$     &          $20.56$ h    &   $+67.4\%$  \\
			Share\&Sync  &  
			$0.9082$ &  $0.9911$ &       $14.97$ h       &  $+22.5\%$     \\
			Share\&Sync$^{\dagger}$  &  
			$0.9084$ &  $0.9901$ &       $12.67$ h       &  $+3.6\%$     \\
			\bottomrule
	\end{tabular}}
	\vspace{-0.5mm}
\end{table}

\noindent \textbf{Effects of Training Manner.} In the above experiments, the teacher and the student are trained synchronously by sharing common input components. Here we  test the effects of such training manner. The results of CTR prediction are shown in Table \ref{effect_of_ctr}.  Training the teacher and the student synchronously achieves almost the same performance as training asynchronously (Ind\&Sync v.s. Ind\&Async). When training the two models by sharing all common input components, the performance of the student degenerates. As the privileged features, i.e., the interacted features between the user and her clicked/purchased items, reflect the user's personal interests, we allocate an independent user id embedding to the student, in order to absorb the extra preferences distilled from the privileged features. The result after such modification is shown in the  row of Share$^*$\&Sync, where the performance degeneration is much alleviated and only a little extra wall-clock time is introduced. We also test the effects of different training manners in the CVR dataset. The results are shown in Table \ref{effect_of_cvr}. As indicated in the rows of Share\&Sync and Ind\&Sync, the teacher can be additionally improved by sharing common input components. Consequently, the student is improved by distilling knowledge from the more accurate teacher.  In the last row of Table \ref{effect_of_cvr}, we report the result of PFD. Compared with the baseline, it  adds almost no extra wall-clock time while achieves similar performance to PFD+MD in the row of Share\&Sync. Overall, by adopting PFD+MD for CTR and PFD for CVR, we can achieve much better performance while add no burden to training time. Therefore,  the technique is even adoptable for online learning on the streaming data \cite{ftrl}, where real-time computation is in high demand.

\section{Conclusion}
In this work, we identify the privileged features existing at Taobao recommendations, i.e., the interacted features for CTR at coarse-grained ranking and the post-event features for CVR at fine-grained ranking. And we propose PFD to leverage them. Different from the traditional LUPI, PFD additionally processes the regular features in the teacher, which is shown to be the core for its success. We also propose PFD+MD to utilize  the complementary feature and model capacities to better instruct the student. And it achieves further improvements. The effectiveness is validated on two fundamental prediction tasks at Taobao recommendations, where the baselines are greatly improved by the proposed distillation techniques. During the on-line A/B tests, the click metric is improved  by $\mathbf{+5.0\%}$ in the CTR task.  And the conversion metric is improved by $\mathbf{+2.3\%}$ in the CVR task. We also address several issues of training PFD, which lead to comparable training speed as the baselines without any distillation.   

\bibliographystyle{ACM-Reference-Format}
\bibliography{feature_distill}


\begin{thebibliography}{40}

%%% the Web address.

\ifx \showCODEN    \undefined \def \showCODEN     #1{\unskip}     \fi
\ifx \showDOI      \undefined \def \showDOI       #1{#1}\fi
\ifx \showISBNx    \undefined \def \showISBNx     #1{\unskip}     \fi
\ifx \showISBNxiii \undefined \def \showISBNxiii  #1{\unskip}     \fi
\ifx \showISSN     \undefined \def \showISSN      #1{\unskip}     \fi
\ifx \showLCCN     \undefined \def \showLCCN      #1{\unskip}     \fi
\ifx \shownote     \undefined \def \shownote      #1{#1}          \fi
\ifx \showarticletitle \undefined \def \showarticletitle #1{#1}   \fi
\ifx \showURL      \undefined \def \showURL       {\relax}        \fi
% invisible to TeX
\providecommand\bibfield[2]{#2}
\providecommand\bibinfo[2]{#2}
\providecommand\natexlab[1]{#1}
\providecommand\showeprint[2][]{arXiv:#2}

\bibitem[\protect\citeauthoryear{Anil, Pereyra, Passos, Ormandi, Dahl, and
  Hinton}{Anil et~al\mbox{.}}{2018}]%
        {dis_distribute}
\bibfield{author}{\bibinfo{person}{Rohan Anil}, \bibinfo{person}{Gabriel
  Pereyra}, \bibinfo{person}{Alexandre Passos}, \bibinfo{person}{Robert
  Ormandi}, \bibinfo{person}{George~E Dahl}, {and} \bibinfo{person}{Geoffrey~E
  Hinton}.} \bibinfo{year}{2018}\natexlab{}.
\newblock \showarticletitle{Large scale distributed neural network training
  through online distillation}. In \bibinfo{booktitle}{\emph{ICLR}}.
\newblock


\bibitem[\protect\citeauthoryear{Ba, Kiros, and Hinton}{Ba
  et~al\mbox{.}}{2016}]%
        {layernorm}
\bibfield{author}{\bibinfo{person}{Jimmy~Lei Ba}, \bibinfo{person}{Jamie~Ryan
  Kiros}, {and} \bibinfo{person}{Geoffrey~E Hinton}.}
  \bibinfo{year}{2016}\natexlab{}.
\newblock \showarticletitle{Layer normalization}.
\newblock \bibinfo{journal}{\emph{arXiv:1607.06450}} (\bibinfo{year}{2016}).
\newblock


\bibitem[\protect\citeauthoryear{Buciluǎ, Caruana, and
  Niculescu-Mizil}{Buciluǎ et~al\mbox{.}}{2006}]%
        {modelcompress}
\bibfield{author}{\bibinfo{person}{Cristian Buciluǎ}, \bibinfo{person}{Rich
  Caruana}, {and} \bibinfo{person}{Alexandru Niculescu-Mizil}.}
  \bibinfo{year}{2006}\natexlab{}.
\newblock \showarticletitle{Model compression}. In
  \bibinfo{booktitle}{\emph{Proceedings of the 12th ACM SIGKDD International
  Conference on Knowledge Discovery and Data Mining}}. ACM,
  \bibinfo{pages}{535--541}.
\newblock


\bibitem[\protect\citeauthoryear{Cheng, Koc, Harmsen, Shaked, Chandra, Aradhye,
  Anderson, Corrado, Chai, Ispir, et~al\mbox{.}}{Cheng et~al\mbox{.}}{2016}]%
        {wide&deep}
\bibfield{author}{\bibinfo{person}{Heng-Tze Cheng}, \bibinfo{person}{Levent
  Koc}, \bibinfo{person}{Jeremiah Harmsen}, \bibinfo{person}{Tal Shaked},
  \bibinfo{person}{Tushar Chandra}, \bibinfo{person}{Hrishi Aradhye},
  \bibinfo{person}{Glen Anderson}, \bibinfo{person}{Greg Corrado},
  \bibinfo{person}{Wei Chai}, \bibinfo{person}{Mustafa Ispir}, {et~al\mbox{.}}}
  \bibinfo{year}{2016}\natexlab{}.
\newblock \showarticletitle{Wide \& deep learning for recommender systems}. In
  \bibinfo{booktitle}{\emph{Proceedings of the 1st workshop on deep learning
  for recommender systems}}. \bibinfo{publisher}{ACM}, \bibinfo{address}{NY,
  USA}, \bibinfo{pages}{7--10}.
\newblock


\bibitem[\protect\citeauthoryear{Covington, Adams, and Sargin}{Covington
  et~al\mbox{.}}{2016}]%
        {youtube}
\bibfield{author}{\bibinfo{person}{Paul Covington}, \bibinfo{person}{Jay
  Adams}, {and} \bibinfo{person}{Emre Sargin}.}
  \bibinfo{year}{2016}\natexlab{}.
\newblock \showarticletitle{Deep neural networks for youtube recommendations}.
  In \bibinfo{booktitle}{\emph{RecSys}}. ACM, \bibinfo{pages}{191--198}.
\newblock


\bibitem[\protect\citeauthoryear{Cybenko}{Cybenko}{1989}]%
        {approximation1}
\bibfield{author}{\bibinfo{person}{George Cybenko}.}
  \bibinfo{year}{1989}\natexlab{}.
\newblock \showarticletitle{Approximation by superpositions of a sigmoidal
  function}.
\newblock \bibinfo{journal}{\emph{Mathematics of Control, Signals and Systems}}
  \bibinfo{volume}{2}, \bibinfo{number}{4} (\bibinfo{year}{1989}),
  \bibinfo{pages}{303--314}.
\newblock


\bibitem[\protect\citeauthoryear{Dean, Corrado, Monga, Chen, Devin, Mao,
  Senior, Tucker, Yang, Le, et~al\mbox{.}}{Dean et~al\mbox{.}}{2012}]%
        {distbelieve}
\bibfield{author}{\bibinfo{person}{Jeffrey Dean}, \bibinfo{person}{Greg
  Corrado}, \bibinfo{person}{Rajat Monga}, \bibinfo{person}{Kai Chen},
  \bibinfo{person}{Matthieu Devin}, \bibinfo{person}{Mark Mao},
  \bibinfo{person}{Andrew Senior}, \bibinfo{person}{Paul Tucker},
  \bibinfo{person}{Ke Yang}, \bibinfo{person}{Quoc~V Le}, {et~al\mbox{.}}}
  \bibinfo{year}{2012}\natexlab{}.
\newblock \showarticletitle{Large scale distributed deep networks}. In
  \bibinfo{booktitle}{\emph{Advances in Neural Information Processing
  Systems}}. \bibinfo{pages}{1223--1231}.
\newblock


\bibitem[\protect\citeauthoryear{Deshpande and Karypis}{Deshpande and
  Karypis}{2004}]%
        {etrac}
\bibfield{author}{\bibinfo{person}{Mukund Deshpande} {and}
  \bibinfo{person}{George Karypis}.} \bibinfo{year}{2004}\natexlab{}.
\newblock \showarticletitle{Item-based top-n recommendation algorithms}.
\newblock \bibinfo{journal}{\emph{ACM Transactions on Information Systems
  (TOIS)}} \bibinfo{volume}{22}, \bibinfo{number}{1} (\bibinfo{year}{2004}),
  \bibinfo{pages}{143--177}.
\newblock


\bibitem[\protect\citeauthoryear{Duchi, Hazan, and Singer}{Duchi
  et~al\mbox{.}}{2011}]%
        {adagrad}
\bibfield{author}{\bibinfo{person}{John Duchi}, \bibinfo{person}{Elad Hazan},
  {and} \bibinfo{person}{Yoram Singer}.} \bibinfo{year}{2011}\natexlab{}.
\newblock \showarticletitle{Adaptive subgradient methods for online learning
  and stochastic optimization}.
\newblock \bibinfo{journal}{\emph{Journal of machine learning research}}
  \bibinfo{volume}{12}, \bibinfo{number}{Jul} (\bibinfo{year}{2011}),
  \bibinfo{pages}{2121--2159}.
\newblock


\bibitem[\protect\citeauthoryear{Guo, Tang, Ye, Li, and He}{Guo
  et~al\mbox{.}}{2017}]%
        {deepfm}
\bibfield{author}{\bibinfo{person}{Huifeng Guo}, \bibinfo{person}{Ruiming
  Tang}, \bibinfo{person}{Yunming Ye}, \bibinfo{person}{Zhenguo Li}, {and}
  \bibinfo{person}{Xiuqiang He}.} \bibinfo{year}{2017}\natexlab{}.
\newblock \showarticletitle{Deep{FM}: {A} factorization-machine based neural
  network for {CTR} prediction}. In \bibinfo{booktitle}{\emph{AAAI}}.
  \bibinfo{publisher}{AAAI Press}, \bibinfo{pages}{1725--1731}.
\newblock


\bibitem[\protect\citeauthoryear{He, Zhang, Ren, and Sun}{He
  et~al\mbox{.}}{2016}]%
        {skipconnect}
\bibfield{author}{\bibinfo{person}{Kaiming He}, \bibinfo{person}{Xiangyu
  Zhang}, \bibinfo{person}{Shaoqing Ren}, {and} \bibinfo{person}{Jian Sun}.}
  \bibinfo{year}{2016}\natexlab{}.
\newblock \showarticletitle{Deep residual learning for image recognition}. In
  \bibinfo{booktitle}{\emph{CVPR}}. \bibinfo{pages}{770--778}.
\newblock


\bibitem[\protect\citeauthoryear{Hidasi, Karatzoglou, Baltrunas, and
  Tikk}{Hidasi et~al\mbox{.}}{2016}]%
        {gru4rec}
\bibfield{author}{\bibinfo{person}{Bal{\'a}zs Hidasi},
  \bibinfo{person}{Alexandros Karatzoglou}, \bibinfo{person}{Linas Baltrunas},
  {and} \bibinfo{person}{Domonkos Tikk}.} \bibinfo{year}{2016}\natexlab{}.
\newblock \showarticletitle{Session-based recommendations with recurrent neural
  networks}. In \bibinfo{booktitle}{\emph{ICLR}}.
\newblock


\bibitem[\protect\citeauthoryear{Hinton, Vinyals, and Dean}{Hinton
  et~al\mbox{.}}{2015}]%
        {distill}
\bibfield{author}{\bibinfo{person}{Geoffrey Hinton}, \bibinfo{person}{Oriol
  Vinyals}, {and} \bibinfo{person}{Jeff Dean}.}
  \bibinfo{year}{2015}\natexlab{}.
\newblock \showarticletitle{Distilling the knowledge in a neural network}.
\newblock \bibinfo{journal}{\emph{arXiv:1503.02531}} (\bibinfo{year}{2015}).
\newblock


\bibitem[\protect\citeauthoryear{Hochreiter and Schmidhuber}{Hochreiter and
  Schmidhuber}{1997}]%
        {lstm}
\bibfield{author}{\bibinfo{person}{Sepp Hochreiter} {and}
  \bibinfo{person}{J{\"u}rgen Schmidhuber}.} \bibinfo{year}{1997}\natexlab{}.
\newblock \showarticletitle{Long short-term memory}.
\newblock \bibinfo{journal}{\emph{Neural computation}} \bibinfo{volume}{9},
  \bibinfo{number}{8} (\bibinfo{year}{1997}), \bibinfo{pages}{1735--1780}.
\newblock


\bibitem[\protect\citeauthoryear{Hornik}{Hornik}{1991}]%
        {approximation2}
\bibfield{author}{\bibinfo{person}{Kurt Hornik}.}
  \bibinfo{year}{1991}\natexlab{}.
\newblock \showarticletitle{Approximation capabilities of multilayer
  feedforward networks}.
\newblock \bibinfo{journal}{\emph{Neural Networks}} \bibinfo{volume}{4},
  \bibinfo{number}{2} (\bibinfo{year}{1991}), \bibinfo{pages}{251--257}.
\newblock


\bibitem[\protect\citeauthoryear{Huang, He, Gao, Deng, Acero, and Heck}{Huang
  et~al\mbox{.}}{2013}]%
        {dssm}
\bibfield{author}{\bibinfo{person}{Po-Sen Huang}, \bibinfo{person}{Xiaodong
  He}, \bibinfo{person}{Jianfeng Gao}, \bibinfo{person}{Li Deng},
  \bibinfo{person}{Alex Acero}, {and} \bibinfo{person}{Larry Heck}.}
  \bibinfo{year}{2013}\natexlab{}.
\newblock \showarticletitle{Learning deep structured semantic models for web
  search using clickthrough data}. In \bibinfo{booktitle}{\emph{CIKM}}. ACM,
  \bibinfo{pages}{2333--2338}.
\newblock


\bibitem[\protect\citeauthoryear{Ioffe and Szegedy}{Ioffe and Szegedy}{2015}]%
        {bn}
\bibfield{author}{\bibinfo{person}{Sergey Ioffe} {and}
  \bibinfo{person}{Christian Szegedy}.} \bibinfo{year}{2015}\natexlab{}.
\newblock \showarticletitle{Batch normalization: Accelerating deep network
  training by reducing internal covariate shift}. In
  \bibinfo{booktitle}{\emph{ICML}}. \bibinfo{pages}{448--456}.
\newblock


\bibitem[\protect\citeauthoryear{Kang and McAuley}{Kang and McAuley}{2018}]%
        {sasrec}
\bibfield{author}{\bibinfo{person}{Wang-Cheng Kang} {and}
  \bibinfo{person}{Julian McAuley}.} \bibinfo{year}{2018}\natexlab{}.
\newblock \showarticletitle{Self-attentive sequential recommendation}. In
  \bibinfo{booktitle}{\emph{ICDM}}. IEEE, \bibinfo{pages}{197--206}.
\newblock


\bibitem[\protect\citeauthoryear{Kim and Rush}{Kim and Rush}{2016}]%
        {dis_translation}
\bibfield{author}{\bibinfo{person}{Yoon Kim} {and}
  \bibinfo{person}{Alexander~M. Rush}.} \bibinfo{year}{2016}\natexlab{}.
\newblock \showarticletitle{Sequence-Level Knowledge Distillation}. In
  \bibinfo{booktitle}{\emph{EMNLP}}. \bibinfo{pages}{1317--1327}.
\newblock


\bibitem[\protect\citeauthoryear{Lambert, Sener, and Savarese}{Lambert
  et~al\mbox{.}}{2018}]%
        {privileged_droput}
\bibfield{author}{\bibinfo{person}{John Lambert}, \bibinfo{person}{Ozan Sener},
  {and} \bibinfo{person}{Silvio Savarese}.} \bibinfo{year}{2018}\natexlab{}.
\newblock \showarticletitle{Deep learning under privileged information using
  heteroscedastic dropout}. In \bibinfo{booktitle}{\emph{CVPR}}.
  \bibinfo{pages}{8886--8895}.
\newblock


\bibitem[\protect\citeauthoryear{Lian, Zhou, Zhang, Chen, Xie, and Sun}{Lian
  et~al\mbox{.}}{2018}]%
        {xdeepfm}
\bibfield{author}{\bibinfo{person}{Jianxun Lian}, \bibinfo{person}{Xiaohuan
  Zhou}, \bibinfo{person}{Fuzheng Zhang}, \bibinfo{person}{Zhongxia Chen},
  \bibinfo{person}{Xing Xie}, {and} \bibinfo{person}{Guangzhong Sun}.}
  \bibinfo{year}{2018}\natexlab{}.
\newblock \showarticletitle{x{D}eep{FM}: {C}ombining explicit and implicit
  feature interactions for recommender systems}. In
  \bibinfo{booktitle}{\emph{Proceedings of the 24th ACM SIGKDD International
  Conference on Knowledge Discovery \& Data Mining}}. \bibinfo{publisher}{ACM},
  \bibinfo{address}{NY, USA}, \bibinfo{pages}{1754--1763}.
\newblock


\bibitem[\protect\citeauthoryear{Lin, Tegmark, and Rolnick}{Lin
  et~al\mbox{.}}{2017}]%
        {why}
\bibfield{author}{\bibinfo{person}{Henry~W Lin}, \bibinfo{person}{Max Tegmark},
  {and} \bibinfo{person}{David Rolnick}.} \bibinfo{year}{2017}\natexlab{}.
\newblock \showarticletitle{Why does deep and cheap learning work so well?}
\newblock \bibinfo{journal}{\emph{Journal of Statistical Physics}}
  \bibinfo{volume}{168}, \bibinfo{number}{6} (\bibinfo{year}{2017}),
  \bibinfo{pages}{1223--1247}.
\newblock


\bibitem[\protect\citeauthoryear{Liu, Xiao, Ou, and Si}{Liu
  et~al\mbox{.}}{2017}]%
        {xinai_cascade}
\bibfield{author}{\bibinfo{person}{Shichen Liu}, \bibinfo{person}{Fei Xiao},
  \bibinfo{person}{Wenwu Ou}, {and} \bibinfo{person}{Luo Si}.}
  \bibinfo{year}{2017}\natexlab{}.
\newblock \showarticletitle{Cascade ranking for operational e-commerce search}.
  In \bibinfo{booktitle}{\emph{Proceedings of the 23rd ACM SIGKDD International
  Conference on Knowledge Discovery and Data Mining}}. ACM,
  \bibinfo{pages}{1557--1565}.
\newblock


\bibitem[\protect\citeauthoryear{Lopez-Paz, Bottou, Sch{\"o}lkopf, and
  Vapnik}{Lopez-Paz et~al\mbox{.}}{2016}]%
        {dis_privileged}
\bibfield{author}{\bibinfo{person}{David Lopez-Paz}, \bibinfo{person}{L{\'e}on
  Bottou}, \bibinfo{person}{Bernhard Sch{\"o}lkopf}, {and}
  \bibinfo{person}{Vladimir Vapnik}.} \bibinfo{year}{2016}\natexlab{}.
\newblock \showarticletitle{Unifying distillation and privileged information}.
  In \bibinfo{booktitle}{\emph{ICLR}}.
\newblock


\bibitem[\protect\citeauthoryear{Maas, Hannun, and Ng}{Maas
  et~al\mbox{.}}{2013}]%
        {leakyrelu}
\bibfield{author}{\bibinfo{person}{Andrew~L Maas}, \bibinfo{person}{Awni~Y
  Hannun}, {and} \bibinfo{person}{Andrew~Y Ng}.}
  \bibinfo{year}{2013}\natexlab{}.
\newblock \showarticletitle{Rectifier nonlinearities improve neural network
  acoustic models}. In \bibinfo{booktitle}{\emph{ICML}},
  Vol.~\bibinfo{volume}{30}. \bibinfo{pages}{3}.
\newblock


\bibitem[\protect\citeauthoryear{McMahan, Holt, Sculley, Young, Ebner, Grady,
  Nie, Phillips, Davydov, Golovin, et~al\mbox{.}}{McMahan
  et~al\mbox{.}}{2013}]%
        {ftrl}
\bibfield{author}{\bibinfo{person}{H~Brendan McMahan}, \bibinfo{person}{Gary
  Holt}, \bibinfo{person}{David Sculley}, \bibinfo{person}{Michael Young},
  \bibinfo{person}{Dietmar Ebner}, \bibinfo{person}{Julian Grady},
  \bibinfo{person}{Lan Nie}, \bibinfo{person}{Todd Phillips},
  \bibinfo{person}{Eugene Davydov}, \bibinfo{person}{Daniel Golovin},
  {et~al\mbox{.}}} \bibinfo{year}{2013}\natexlab{}.
\newblock \showarticletitle{ftrl}. In \bibinfo{booktitle}{\emph{Proceedings of
  the 19th ACM SIGKDD international conference on Knowledge discovery and data
  mining}}. \bibinfo{pages}{1222--1230}.
\newblock


\bibitem[\protect\citeauthoryear{Mishra and Marr}{Mishra and Marr}{2018}]%
        {dis_quant}
\bibfield{author}{\bibinfo{person}{Asit Mishra} {and} \bibinfo{person}{Debbie
  Marr}.} \bibinfo{year}{2018}\natexlab{}.
\newblock \showarticletitle{Apprentice: Using knowledge distillation techniques
  to improve low-precision network accuracy}. In
  \bibinfo{booktitle}{\emph{ICLR}}.
\newblock


\bibitem[\protect\citeauthoryear{Ni, Ou, Liu, Li, Ou, Zeng, and Si}{Ni
  et~al\mbox{.}}{2018}]%
        {dupn}
\bibfield{author}{\bibinfo{person}{Yabo Ni}, \bibinfo{person}{Dan Ou},
  \bibinfo{person}{Shichen Liu}, \bibinfo{person}{Xiang Li},
  \bibinfo{person}{Wenwu Ou}, \bibinfo{person}{Anxiang Zeng}, {and}
  \bibinfo{person}{Luo Si}.} \bibinfo{year}{2018}\natexlab{}.
\newblock \showarticletitle{Perceive your users in depth: Learning universal
  user representations from multiple e-commerce tasks}. In
  \bibinfo{booktitle}{\emph{Proceedings of the 24th ACM SIGKDD International
  Conference on Knowledge Discovery \& Data Mining}}. ACM,
  \bibinfo{pages}{596--605}.
\newblock


\bibitem[\protect\citeauthoryear{Pereyra, Tucker, Chorowski, Kaiser, and
  Hinton}{Pereyra et~al\mbox{.}}{2017}]%
        {regularize_distill}
\bibfield{author}{\bibinfo{person}{Gabriel Pereyra}, \bibinfo{person}{George
  Tucker}, \bibinfo{person}{Jan Chorowski}, \bibinfo{person}{{\L}ukasz Kaiser},
  {and} \bibinfo{person}{Geoffrey Hinton}.} \bibinfo{year}{2017}\natexlab{}.
\newblock \showarticletitle{Regularizing neural networks by penalizing
  confident output distributions}.
\newblock \bibinfo{journal}{\emph{arXiv:1701.06548}} (\bibinfo{year}{2017}).
\newblock


\bibitem[\protect\citeauthoryear{Romero, Ballas, Kahou, Chassang, Gatta, and
  Bengio}{Romero et~al\mbox{.}}{2015}]%
        {dis_featuremap}
\bibfield{author}{\bibinfo{person}{Adriana Romero}, \bibinfo{person}{Nicolas
  Ballas}, \bibinfo{person}{Samira~Ebrahimi Kahou}, \bibinfo{person}{Antoine
  Chassang}, \bibinfo{person}{Carlo Gatta}, {and} \bibinfo{person}{Yoshua
  Bengio}.} \bibinfo{year}{2015}\natexlab{}.
\newblock \showarticletitle{Fitnets: Hints for thin deep nets}. In
  \bibinfo{booktitle}{\emph{ICLR}}.
\newblock


\bibitem[\protect\citeauthoryear{Ruder}{Ruder}{2017}]%
        {mtl}
\bibfield{author}{\bibinfo{person}{Sebastian Ruder}.}
  \bibinfo{year}{2017}\natexlab{}.
\newblock \showarticletitle{An overview of multi-task learning in deep neural
  networks}.
\newblock \bibinfo{journal}{\emph{arXiv:1706.05098}} (\bibinfo{year}{2017}).
\newblock


\bibitem[\protect\citeauthoryear{Szegedy, Vanhoucke, Ioffe, Shlens, and
  Wojna}{Szegedy et~al\mbox{.}}{2016}]%
        {noise}
\bibfield{author}{\bibinfo{person}{Christian Szegedy}, \bibinfo{person}{Vincent
  Vanhoucke}, \bibinfo{person}{Sergey Ioffe}, \bibinfo{person}{Jon Shlens},
  {and} \bibinfo{person}{Zbigniew Wojna}.} \bibinfo{year}{2016}\natexlab{}.
\newblock \showarticletitle{Rethinking the inception architecture for computer
  vision}. In \bibinfo{booktitle}{\emph{CVPR}}. \bibinfo{pages}{2818--2826}.
\newblock


\bibitem[\protect\citeauthoryear{Tang and Wang}{Tang and Wang}{2018}]%
        {dis_rank}
\bibfield{author}{\bibinfo{person}{Jiaxi Tang} {and} \bibinfo{person}{Ke
  Wang}.} \bibinfo{year}{2018}\natexlab{}.
\newblock \showarticletitle{Ranking distillation: Learning compact ranking
  models with high performance for recommender system}. In
  \bibinfo{booktitle}{\emph{Proceedings of the 24th ACM SIGKDD International
  Conference on Knowledge Discovery \& Data Mining}}. ACM,
  \bibinfo{pages}{2289--2298}.
\newblock


\bibitem[\protect\citeauthoryear{Vapnik and Izmailov}{Vapnik and
  Izmailov}{2015}]%
        {privilegedinformationlearning}
\bibfield{author}{\bibinfo{person}{Vladimir Vapnik} {and} \bibinfo{person}{Rauf
  Izmailov}.} \bibinfo{year}{2015}\natexlab{}.
\newblock \showarticletitle{Learning using privileged information: similarity
  control and knowledge transfer.}
\newblock \bibinfo{journal}{\emph{Journal of Machine Learning Research}}
  \bibinfo{volume}{16}, \bibinfo{number}{2023-2049} (\bibinfo{year}{2015}),
  \bibinfo{pages}{2}.
\newblock


\bibitem[\protect\citeauthoryear{Vapnik and Vashist}{Vapnik and
  Vashist}{2009}]%
        {A_new_learning}
\bibfield{author}{\bibinfo{person}{Vladimir Vapnik} {and}
  \bibinfo{person}{Akshay Vashist}.} \bibinfo{year}{2009}\natexlab{}.
\newblock \showarticletitle{A new learning paradigm: Learning using privileged
  information}.
\newblock \bibinfo{journal}{\emph{Neural Networks}} \bibinfo{volume}{22},
  \bibinfo{number}{5-6} (\bibinfo{year}{2009}), \bibinfo{pages}{544--557}.
\newblock


\bibitem[\protect\citeauthoryear{Vaswani, Shazeer, Parmar, Uszkoreit, Jones,
  Gomez, Kaiser, and Polosukhin}{Vaswani et~al\mbox{.}}{2017}]%
        {attentionall}
\bibfield{author}{\bibinfo{person}{Ashish Vaswani}, \bibinfo{person}{Noam
  Shazeer}, \bibinfo{person}{Niki Parmar}, \bibinfo{person}{Jakob Uszkoreit},
  \bibinfo{person}{Llion Jones}, \bibinfo{person}{Aidan~N Gomez},
  \bibinfo{person}{{\L}ukasz Kaiser}, {and} \bibinfo{person}{Illia
  Polosukhin}.} \bibinfo{year}{2017}\natexlab{}.
\newblock \showarticletitle{Attention is all you need}. In
  \bibinfo{booktitle}{\emph{Advances in Neural Information Processing
  Systems}}. \bibinfo{pages}{5998--6008}.
\newblock


\bibitem[\protect\citeauthoryear{Wang, Zhang, Sun, and Qi}{Wang
  et~al\mbox{.}}{2018}]%
        {kdgan}
\bibfield{author}{\bibinfo{person}{Xiaojie Wang}, \bibinfo{person}{Rui Zhang},
  \bibinfo{person}{Yu Sun}, {and} \bibinfo{person}{Jianzhong Qi}.}
  \bibinfo{year}{2018}\natexlab{}.
\newblock \showarticletitle{Kdgan: Knowledge distillation with generative
  adversarial networks}. In \bibinfo{booktitle}{\emph{Advances in Neural
  Information Processing Systems}}. \bibinfo{pages}{775--786}.
\newblock


\bibitem[\protect\citeauthoryear{Zhang, Xiang, Hospedales, and Lu}{Zhang
  et~al\mbox{.}}{2018}]%
        {dis_mutulearn}
\bibfield{author}{\bibinfo{person}{Ying Zhang}, \bibinfo{person}{Tao Xiang},
  \bibinfo{person}{Timothy~M Hospedales}, {and} \bibinfo{person}{Huchuan Lu}.}
  \bibinfo{year}{2018}\natexlab{}.
\newblock \showarticletitle{Deep mutual learning}. In
  \bibinfo{booktitle}{\emph{CVPR}}. \bibinfo{pages}{4320--4328}.
\newblock


\bibitem[\protect\citeauthoryear{Zhou, Fan, Cui, Bian, Zhu, and Gai}{Zhou
  et~al\mbox{.}}{2018a}]%
        {rocket}
\bibfield{author}{\bibinfo{person}{Guorui Zhou}, \bibinfo{person}{Ying Fan},
  \bibinfo{person}{Runpeng Cui}, \bibinfo{person}{Weijie Bian},
  \bibinfo{person}{Xiaoqiang Zhu}, {and} \bibinfo{person}{Kun Gai}.}
  \bibinfo{year}{2018}\natexlab{a}.
\newblock \showarticletitle{Rocket launching: A universal and efficient
  framework for training well-performing light net}. In
  \bibinfo{booktitle}{\emph{AAAI}}.
\newblock


\bibitem[\protect\citeauthoryear{Zhou, Zhu, Song, Fan, Zhu, Ma, Yan, Jin, Li,
  and Gai}{Zhou et~al\mbox{.}}{2018b}]%
        {din}
\bibfield{author}{\bibinfo{person}{Guorui Zhou}, \bibinfo{person}{Xiaoqiang
  Zhu}, \bibinfo{person}{Chenru Song}, \bibinfo{person}{Ying Fan},
  \bibinfo{person}{Han Zhu}, \bibinfo{person}{Xiao Ma},
  \bibinfo{person}{Yanghui Yan}, \bibinfo{person}{Junqi Jin},
  \bibinfo{person}{Han Li}, {and} \bibinfo{person}{Kun Gai}.}
  \bibinfo{year}{2018}\natexlab{b}.
\newblock \showarticletitle{Deep interest network for click-through rate
  prediction}. In \bibinfo{booktitle}{\emph{Proceedings of the 24th ACM SIGKDD
  International Conference on Knowledge Discovery \& Data Mining}}. ACM,
  \bibinfo{pages}{1059--1068}.
\newblock


\end{thebibliography}
\end{document}